\documentclass[12pt]{article}

\usepackage{epsfig}
\usepackage{tabularx}
\usepackage{graphicx}
\usepackage{amssymb,amsfonts,amsmath}

\begin{document}

\begin{titlepage}



\centerline{\large \bf {Statefinder parameters in two dark energy models}}

\vskip 1cm

\centerline{Grigoris Panotopoulos}

\vskip 1cm

\centerline{ASC, Physics Department LMU,}

\vskip 0.2 cm

\centerline{Theresienstr. 37, 80333 Munich, Germany}

\vskip 0.2 cm

\centerline{email:{\it Grigoris.Panotopoulos@physik.uni-muenchen.de}}

\begin{abstract}
The statefinder parameters ($r,s$) in two dark energy models are studied. In the first, we discuss in four-dimensional General Relativity a two fluid model, in which dark energy and dark matter are allowed to interact with each other. In the second model, we consider the DGP brane model generalized by taking a possible energy exchange between the brane and the bulk into account. We determine the values of the statefinder parameters that correspond to the unique attractor of the system at hand. Furthermore, we produce plots in which we show $s,r$ as functions of red-shift, and the ($s-r$) plane for each model.
\end{abstract}

\end{titlepage}

\section{Introduction}

A plethora of observational data are now available, which show that we live in a
flat universe that expands with an accelerating rate and that the dominant component in the energy budget of the universe
is an unusual material, the nature of which still remains unknown. Identifying the origin and nature of dark energy is
one of the great challenges in modern theoretical cosmology. The simplest candidate for dark energy is the cosmological
constant, which corresponds to a perfect fluid with state parameter $w=p/\rho=-1$. The LCDM model is still in agreement with all observational data. However, due to the problems associated to the cosmological constant, over the years many other
theoretical models have been proposed and studied. One class of such models is based on some modification of Einstein's
gravity~\cite{gde1,gde2} and one is talking about the so-called geometrical dark energy models. Another class contains the dynamical dark
energy models, in which a new dynamical field (almost certainly a scalar field) is coupled to gravity. In the second
class one would find models called quintessence~\cite{quintessence}, phantom~\cite{phantom}, quintom~\cite{quintom}, k-essence~\cite{kessence}, tachyonic~\cite{tachyonic} etc. A recent review on dark energy dynamics one can find in~\cite{copeland}. Furthermore, an attempt to solve, or at least to alleviate, the coincidence problem is realized in models of interacting dark energy~\cite{interacting}. In this class of models an interaction between the dark matter and dark energy is allowed, which after all is a natural possibility to be considered.

Nowadays several cosmological models of dark energy are available which cannot be excluded by current observational data. However, a few years ago two new cosmological parameters were introduced~\cite{sahni} in order to discriminate between different dark energy models. These are the so called statefinder parameters and are given entirely in terms of the scale factor and its derivatives with respect to the cosmic time, up to the third order. The statefinder parameters are defined as follows
\begin{eqnarray}
r & = & \frac{\dddot{a}}{a H^3} \\
s & = & \frac{r-1}{3 (q-\frac{1}{2})}
\end{eqnarray}
where $a$ is the scale factor of the universe, $H$ is the Hubble parameter, a dot denotes differentiation with respect to the cosmic time $t$ and $q$ is the deceleration parameter
\begin{equation}
q=-\frac{\ddot{a}}{a H^2}
\end{equation}
After the deceleration parameter with the second derivative of the scale factor with respect to the cosmic time, a natural quantity to be considered is $r$ with the third derivative of $a$ with respect to $t$. For the LCDM model, $r=1$ and $s=0$. At this point the easiest way to see this is to consider the evolution of the universe at large times, that is when the contribution from matter is negligible. Then the universe expands due to the cosmological constant only and the scale factor grows exponentially with time, $a(t)=\textrm{exp}(t)$. Then one can immediately see that $r=1$ and therefore $s=0$ (since $q$=-1). 

The trajectories in the $r-s$ plane for various existing models can exhibit quite different behaviors. The deviation of these trajectories from the $(0,1)$ point defines the distance of a given model from the LCDM model. The statefinder pair ($r,s$) can successfully differentiate between a wide variety of cosmological models including a cosmological constant, brane models, quintessence, Chaplygin gas, and interacting dark energy models. In a given model the pair ${r,s}$ can be computed and the trajectory in the $r-s$ plane can be drown. Furthermore, the values of $r,s$ can be extracted from future observations~\cite{snap}. Therefore, the statefinder diagnostic combined with future observations may possibly be used to discriminate between different dark energy models.

Up to now, the statefinder diagnostic has been applied to several models, see e.g.~\cite{statefinder}. In the present work we wish to study two dark energy models. In the first we consider in four-dimensional General Relativity a two fluid model with dark matter and dark energy interacting with each other. The form of the interaction is specified below. We do not rely on a concrete particle physics model for dark energy. We just treat dark energy as a hydrodynamical fluid with a constant state parameter $w=p_X/\rho_X$, where $p_X, \rho_X$ are the pressure and energy density of dark energy respectively. In the second model, we consider the DGP brane model generalized by taking into account a possible energy exchange between the brane and the bulk.

\section{The dark energy models}

\subsection{Interacting model}

The equations of motion for our system are Friedmann equations and the semi-conservation equation for each fluid component
\begin{eqnarray}
H^2 & = & \frac{\kappa^2}{3} \: \rho \\
\dot{H} & = & -\frac{\kappa^2}{2} \: (\rho+p) \\
Q & = & \dot{\rho}_m+3H \rho_m \\
-Q & = & \dot{\rho}_X+3H (\rho_X+p_X)
\end{eqnarray}
where $Q$ is a source term responsible for the interaction between the two fluid components, $\kappa^2=8 \pi G$, $\rho_m$ is the energy density of matter ($p_m=0$), $\rho=\rho_m+\rho_X$ is the total energy density and $p=p_m+p_X=p_X$ is the total pressure. The first Friedmann equation is a constraint, while the second one is a dynamical equation.
Below we shall assume that the interaction term $Q$ takes the form
\begin{equation}
Q=\delta H \rho_m
\end{equation}
where $\delta$ is a dimensionless quantity which for simplicity is taken to be a constant. As a matter of fact the present model with the given form of the interaction  term was very recently compared to observational data~\cite{japans}. In that work it was found that the allowed range for $\delta, w, \Omega_{X,0}$ is the following
\begin{equation}
-0.08 < \delta < 0.03
\end{equation}
\begin{equation}
-1.16 < w < -0.91
\end{equation}
\begin{equation}
0.69 < \Omega_{X,0} < 0.77
\end{equation}
while the best-fit parameters were found to be
\begin{eqnarray}
\delta & = & -0.03 \\
w & = & -1.02 \\
\Omega_{X,0} & = & 0.73
\end{eqnarray}
where $\Omega_{X,0}$ is the present value of the normalized density of dark energy. The normalized densities for matter and dark energy are defined by
\begin{eqnarray}
\Omega_m=\frac{\kappa^2 \rho_m}{3H^2} \\
\Omega_X=\frac{\kappa^2 \rho_X}{3H^2}
\end{eqnarray}
and each of them $0 \leq \Omega_i \leq 1$. Then the first Friedmann equation takes the form
\begin{equation}
\Omega_m+\Omega_X = 1
\end{equation}
The second Friedmann equation is written in terms of the normalized densities
\begin{equation}
\dot{H}=-\frac{3}{2} H^2 (\Omega_m+\gamma \Omega_X)
\end{equation}
where $\gamma=1+w$. If we define $N=\textrm{ln}a$ we can write down the equation of motion for $\Omega_m$ or $\Omega_X$ with respect to $N$. One finds
\begin{equation}
\Omega_X'=-(1-\Omega_X) (\delta+3w \Omega_X)
\end{equation}
and
\begin{equation}
\Omega_m'=(1-\Omega_X) (\delta+3w \Omega_X)
\end{equation}
where a prime denotes differentiation with respect to $N$. We see that $\Omega_m'=-\Omega_X'$, as it should since $\Omega_X+\Omega_m=1$. Since only one of the normalized densities is independent, we shall consider the equation of motion for $\Omega_X$ and for simplicity we shall drop the index $X$ below. It is obvious that there are two critical points, namely
\begin{eqnarray}
\Omega_{*,1} & = & 1 \\
\Omega_{*,2} & = & -\frac{\delta}{3 w}
\end{eqnarray}
To determine the stability of the critical points we have to linearize the system. For the first critical point we set $\Omega=1+\delta \Omega$ and we obtain
\begin{equation}
\delta \Omega'=(\delta+3w) \delta \Omega
\end{equation}
which means that the critical point is stable for $\delta+3w < 0$ and unstable for $\delta+3w > 0$. In fact according to the values obtained observationally the critical point $\Omega_{*,1}=1$ is stable. For the second critical point $\Omega_{*,2}=-\delta/(3w)$ we set $\Omega=-\delta/(3w)+\delta \Omega$ and we obtain
\begin{equation}
\delta \Omega'=-3w(1-\Omega_{*,2}) \delta \Omega
\end{equation}
which means that it is unstable.

Up to now we have determined the number and stability of the critical points of the system. Now we shall focus to the stable critical point and obtain the values of $r,s$ corresponding to that. Using the equations of motion and the definition for the statefinder parameters we obtain
\begin{eqnarray}
s & = & 1+w+\frac{\delta}{3} (\frac{1}{\Omega_X}-1) \\
r & = & 1+\frac{9}{2} \: w \Omega_X s
\end{eqnarray}
Notice that for the LCDM model, for which $\delta=0$ and $w=-1$, one obtains that $s=0, r=1$ at every instant of time and not only at large times.
At the critical point $\Omega_{X,*}=1$ their values are given by
\begin{eqnarray}
s & = & 1+w \\
r & = & 1+\frac{9}{2}w (1+w)
\end{eqnarray}
Notice that the dependence on the interaction $\delta$ drops out and the values of the statefinder parameters at the stable critical point only depend on the dark energy state parameter $w$. For a $w \simeq -1$, $s \simeq 0$ and $r \simeq 1$, that is the point ($s,r$) is only slightly different from that corresponding to LCDM. The ($s-r$) plane for this model is shown in Fig.~1, in which the different orbits correspond to different values of $\delta=-0.08,-0.03,0.03$. For the numerical demands we have let $N=\textrm{ln}a$ go up to $N=10$, which is sufficient for our purposes. The statefinders $r,s$ as functions of red-shift are shown in Fig.~3 and Fig.~4. The present values of $r,s$ are as follows

\begin{center}
\begin{tabular}{|c|c|c|}
\hline
 $w=-1.02$, $\delta=-0.08$ & $s(0)=-0.030$ & $r(0)=1.100$ \\  \hline
 $w=-1.02$, $\delta=-0.03$ & $s(0)=-0.024$ & $r(0)=1.079$ \\  \hline
 $w=-1.02$, $\delta=0.03$ & $s(0)=-0.016$ & $r(0)=1.055$ \\ \hline
\multicolumn{3}{l}{Table 1: The present values of $r,s$ for the four-dimensional interacting dark energy model.}
\end{tabular}
\end{center}
Finally, in Fig.~5 we show as a comparison where the curves corresponding to quintessence and constant $w$ without an interaction are located on the ($s-r$) plane.

\subsection{Brane model}

We shall now discuss another dark energy model, this time a brane model. In particular we wish to study the DGP model~\cite{dgp} taking into account the energy exchange between the brane and the bulk~\cite{gde2}c. Let us first review the basic formulae that we shall be using. The model is defined by the action
\begin{equation}
S=\int \!d^{5}x\sqrt{-g}\,(M^{3}R-\Lambda)\,+\int
\!d^{4}x\sqrt{-h}\,(m^{2}\hat{R}-V)
\end{equation}
plus a matter content both in the bulk and on the brane, where $R, \hat{R}$ are the Ricci scalars of the bulk metric $g_{AB}$ and the induced metric $h_{\mu \nu}$ respectively.
The five-dimensional Planck mass is $M$, the bulk cosmological constant is $\Lambda/2M^{3}\!<\!0$, the brane tension is $V$, and the
induced-gravity crossover scale is $r_{c}\!=\!m^{2}/M^{3}$. Assuming a flat universe and a perfect fluid on the brane with state parameter $w$, the cosmological equations on the brane read
\begin{eqnarray}
&&\,\,\,\,\,\,\,\,\,\,\,\,\dot{\rho}+3(1+w)H\rho=-T \\
&&\,\,\,\,\,\,\,\,\,\,\,H^{2}=\mu+2\gamma \rho + \beta
\psi
\label{eq:31}\\
&&\!\!\!\!\!\!
\dot{\psi}+2H\Big(\!\psi-\frac{\lambda+6(1\!-\!3w)\gamma
\rho}{\psi}\!\Big)=\frac{2\gamma T}{\beta}
\label{eq:35}\\
&&\!\!\!\!\!\!\!\frac{\ddot{a}}{a}=\mu-(1\!+\!3w)\gamma \rho +
\beta \frac{\lambda+6(1\!-\!3w)\gamma \rho}{\psi} \label{eq:32}
\end{eqnarray}
where $T$ is the term responsible for the energy exchange between the brane and the bulk and it is assumed to have the form $T=A \rho^{\nu}$~\cite{exchange} with $A,\nu$ constants. The case $A >0$ corresponds to outflow while the case $A < 0$ corresponds to influx. The new parameters $\beta, \gamma, \lambda, \mu$ are related to the old ones $M,m,V,\Lambda$ by
\begin{eqnarray}
\lambda & = & \frac{2V}{m^{2}}+\frac{12}{r_{c}^{2}}-\frac{\Lambda}{M^{3}} \\
\mu & = & \frac{V}{6m^{2}}+\frac{2}{r_{c}^{2}} \\
\gamma & = & \frac{1}{12m^{2}} \\
\beta & = & \frac{1}{\sqrt{3}r_{c}}
\end{eqnarray}
and adopting the Randall-Sundrum condition, $\mu=-\beta \sqrt{\lambda}$. The critical point analysis has be done in Ref.~\cite{gde2}c and in the case of dust and influx the results can be shown in the table below
\begin{center}
\begin{tabular}{|c|c|c|c|}
\hline
 & $\nu<3/2$ & $\nu=3/2$ & $\nu>3/2$  \\ \hline
 No. of F.P. & 1 & 0 or 1 & 1 \\ \hline
 Nature & \textsf{A} & \,\,\,\,\,\,\,\,\,\,\,\,\textsf{A} & \textsf{S} \\
\hline
\multicolumn{4}{l}{Table 2: The fixed points for w=0, influx}
\end{tabular}
\end{center}
where the second row shows the number of critical points, while the third row shows the nature of the fixed points, attractor (A) or saddle (S). Below we shall be interested in the $\nu < 3/2$ case, since this is when the system possesses always a unique attractor. For the critical point analysis it is useful to define appropriate dimensionless quantities. Defining
\begin{equation} \omega_{m}\!=\!\frac{2\gamma\rho}{D^{2}}
\,\,\,\,\,,\,\,\,\,\,\omega_{\psi}=\frac{\beta\psi}{D^{2}}\,\,\,\,\,,\,\,\,\,\,
Z=\frac{H}{D}\, \label{flat}
\end{equation}
where
$D\!=\!\sqrt{H^{2}\!-\!\mu}$, we obtain the equations
\begin{eqnarray}
&&\,\,\,\,\,\,\,\,\,\,\,\,\,\,\,\,\,\,\,\,\,\,\,\,\,\,\,\,\,\,\,\,\,\,\,\,\,\,\,\,\,\,\,
\omega_{m}+\omega_{\psi}=1
\label{friflat}\\
&&\!\!\!\!\!\!\!\omega_{m}'\!=\!\omega_{\!m}\!\Big[\!(1\!+\!3w)(\omega_{\!m}\!\!-\!1\!)Z\!-\!\frac{
A}{\sqrt{|\mu|}}
\Big(\!\frac{|\mu|\omega_{\!m}}{2\gamma}\!\Big)^{\!\!\nu\!-\!1}\!(1\!-\!Z^{2})^{\frac{3}{2}-\nu}\nonumber\\
&&\,\,\,\,\,\,\,\,\,\,\,\,\,\,\,\,\,\,\,\,\,\,\,
-2Z(1\!-\!Z^{2})\frac{1\!-\!Z^{2}\!-\!3(1\!-\!3w)\beta^{2}\mu^{-1}\omega_{m}}
{1\!-\!\omega_{m}}\!\Big] \label{gerold}\\
&&\!\!\!\!\!\!\!Z'\!=\!(1\!-\!Z^{2})\Big[(1\!-\!Z^{2})
\frac{1\!-\!Z^{2}\!-\!3(1\!-\!3w)\beta^{2}\mu^{-1}\omega_{m}}{1\!-\!\omega_{m}}-1\nonumber\\
&& \,\,\,\,\,\,\,\,\,\,\,\,\,\,\,\,\,\,\,\,\,\,\,\,
\,\,\,\,\,\,\,\,\,\,\,\,\,\,\,\,\,\,\,\,\,\,\,\,\,\,\,\,\,\,\,\,\,\,\,\,\,\,\,\,\,\,\,\,\,\,\,\,
\,\,\,\,\,\,\,\,\,\,\,\,\,\,\,\,\,\,\,\,-\frac{1\!+\!3w}{2}\omega_{m}\Big]
\label{italy}
\end{eqnarray}
with $'\!=\!d/d\tau\!=\!D^{-1}d/dt$, while the
deceleration parameter is given by
\begin{equation}
q\!=\!\frac{1}{Z^{2}}\!\Big[\!\frac{1\!+\!3w}{2}\omega_{m}\!-\!(1\!-\!Z^{2})
\frac{\omega_{m}\!-\!\!Z^{2}\!-\!3(1\!-\!3w)\beta^{2}\mu^{-1}\omega_{m}}{1\!-\!
\omega_{m}}\!\Big]\label{greece}
\end{equation}
Now we shall determine the statefinder parameters in terms of $Z, \omega_m$ using their definition and the cosmological equations. The expressions are lengthy, therefore we choose to express them in term of the deceleration parameter
\begin{eqnarray}
r & = & q+2q^2-\frac{q'}{Z} \\
s & = & \frac{r-1}{3 (q-\frac{1}{2})}
\end{eqnarray}
One understands that by taking the derivative of $q$ with respect to $\tau$ and using the equations of motion for $Z, \omega_m$, it is possible to express $r,s$ in terms of the latter variables. In practice, we solve the system numerically and obtain $Z, \omega_m$ as functions of $\tau$. Then we compute $q, q'$ and finally the statefinder parameters $\omega_m(\tau), Z(\tau)$. At the critical points, $q'=0$ and in this particular model it turns out~\cite{gde2}c that $q_*=-1$ always holds. Therefore we find that
\begin{eqnarray}
r_* & = & 1 \\
s_* & = & 0
\end{eqnarray}
at the critical points. These are the values corresponding to the LCDM model. The ($s-r$) plane for this model is shown in Fig.~2, in which the different orbits correspond to different values of $\nu=1,1.1,1.2$. For our computational needs we have allowed the mathematical time $\tau$ go up to $\tau=100$, although a few tens would also be sufficient. The basic idea pursued in~\cite{gde2}c was that our universe is close to the unique attractor of the system studied. Therefore today's values of $r,s$ are approximately the ones corresponding to the fixed point. Contrary to the previous four-dimensional interacting model, we cannot show for a comparison in the same ($s-r$) plot the curves corresponding to the brane model and to quintessence and constant $w$ without an interaction. The reason is because the curves for quintessence and constant $w$ are located close to the cosmological constant point ($0,1$) and cannot be seen in the plot. Finally, in Fig.~6,7 we show the statefinder parameters $s,r$ as functions of red-shift $z$ for three different values of $\nu=1,1.1,1.2$.

\section{Conclusions}

To summarize our work, we have considered two dark energy models. In the first model we have considered in four-dimensional General Relativity a two fluid model in which an interaction is allowed between dark energy and dark matter. We have treated dark energy as a hydrodynamical fluid with a constant state parameter $w$ and we have not relied on a concrete particle physics model for dark energy. Using dynamical system methods we have determined the number and stability of the critical points of the system. We have found that there is a stable and an unstable critical point and have computed the values of the pair ($s,r$) corresponding to the stable critical point. It turns out that these values depend on the dark energy state parameter $w$ only and that the dependence on the interaction $\delta$ drops out. Our conclusion is that for the observational value of $w \simeq -1$, the values of ($s,r$) for the system under study is only slightly different from that of LCDM. In the other model we have considered the DGP brane model taking into account a possible energy exchange between the brane and the bulk. The assumed form for this energy exchange as well as the number and nature of the critical points for this model have been previously discussed. Here we have determined the values of the statefinder parameters corresponding to the unique attractor of the model and we have shown that are identical to the values corresponding to the LCDM model. Finally, we have generated plots in which we show $s,r$ as functions of red-shift, and the ($s-r$) plane for each model.

\section*{Acknowlegements}

We would like to thank the anonymous reviewer for his/her valuable comments and suggestions that greatly improved the quality of the presentation. This work was supported by project "Particle Cosmology".

\newpage

\begin{figure}
\centerline{\epsfig{figure=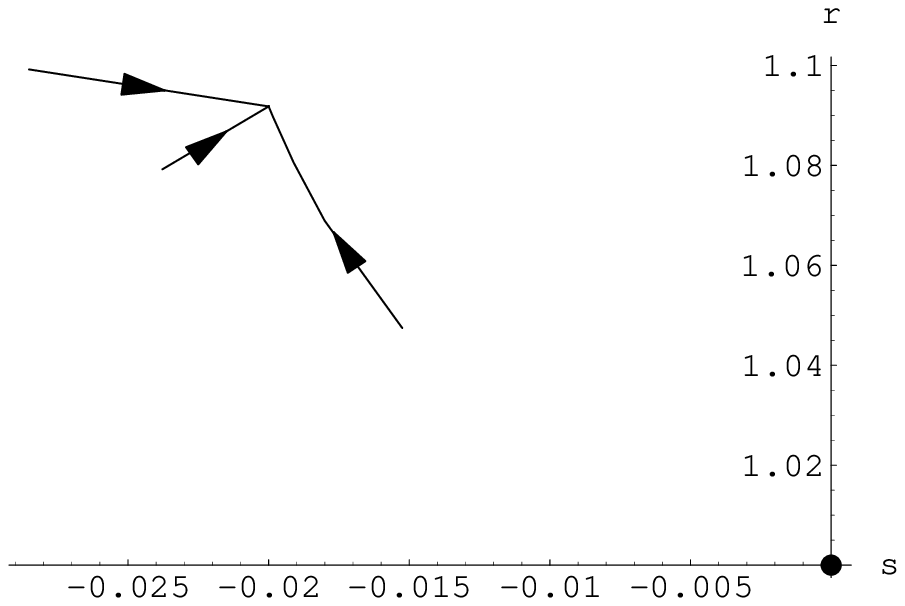,height=6cm,angle=0}}
\caption{The ($s-r$) plane for the four-dimensional interacting dark energy model. The different orbits correspond to different values of $\delta=-0.08,-0.03,0.03$. The marked point at (0,1) corresponds to the LCDM model.}
\end{figure}

\begin{figure}
\centerline{\epsfig{figure=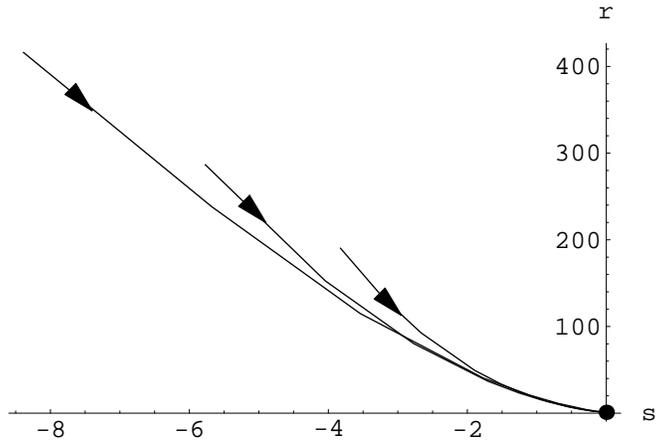,height=6cm,angle=0}}
\caption{The ($s-r$) plane for the DGP brane model. The different orbits correspond to different values of $\nu=1,1.1,1.2$. The marked point at (0,1) corresponds to the LCDM model.}
\end{figure}

\begin{figure}
\centerline{\epsfig{figure=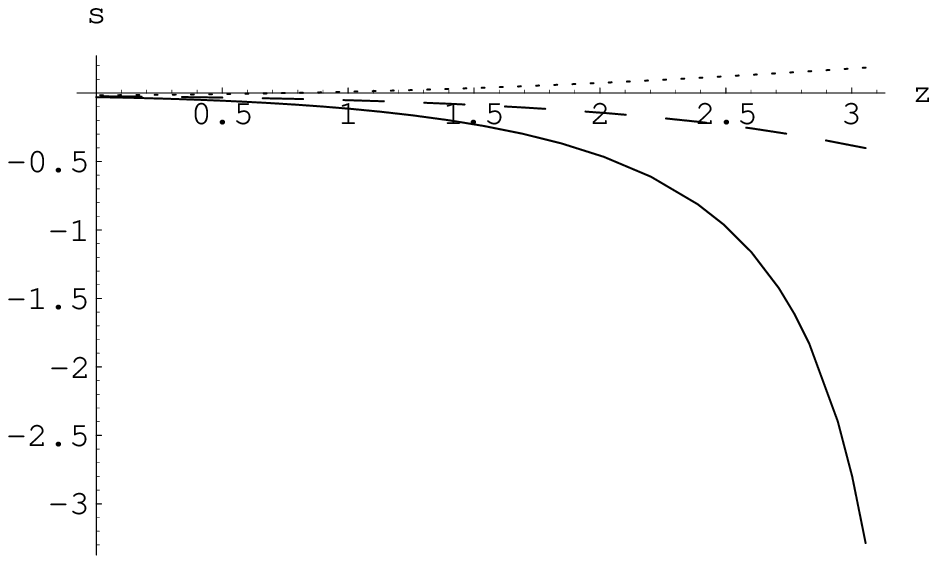,height=6cm,angle=0}}
\caption{Parameter $s$ as a function of red-shift $z$ for the interacting model. The solid line corresponds to $\delta=-0.08$, the dashed line corresponds to $\delta=-0.03$ and the dotted line corresponds to $\delta=0.03$.}
\end{figure}

\begin{figure}
\centerline{\epsfig{figure=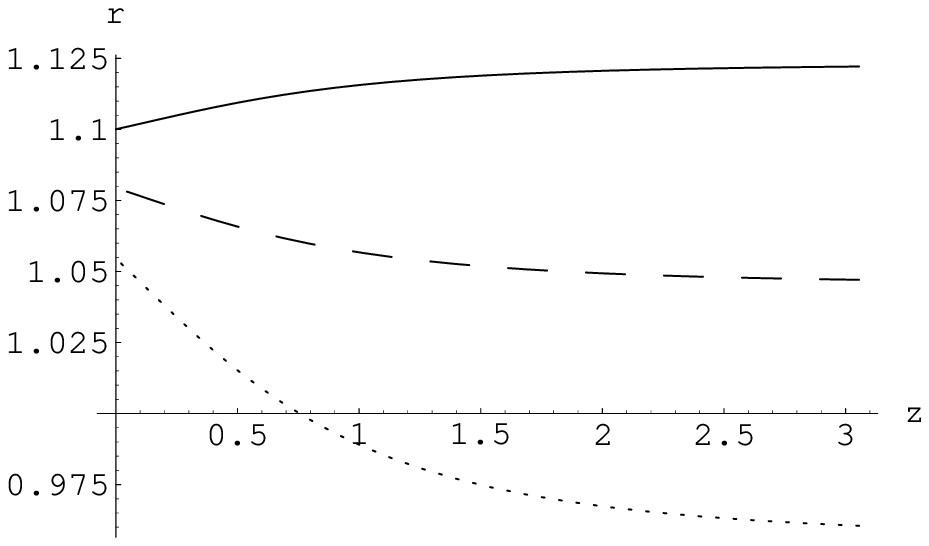,height=6cm,angle=0}}
\caption{Parameter $r$ as a function of red-shift $z$ for the interacting model. The solid line corresponds to $\delta=-0.08$, the dashed line corresponds to $\delta=-0.03$ and the dotted line corresponds to $\delta=0.03$.}
\end{figure}

\begin{figure}
\centerline{\epsfig{figure=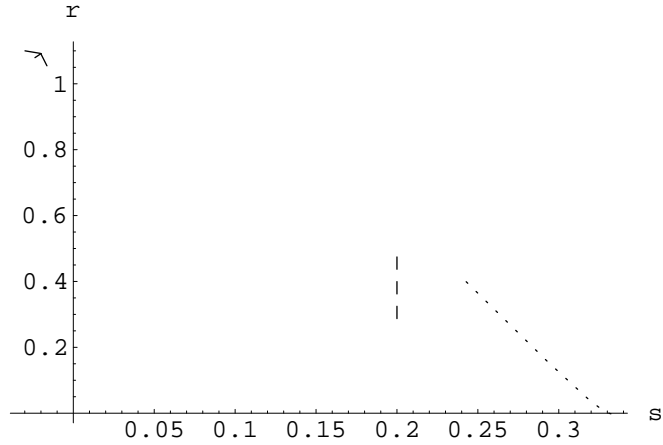,height=6cm,angle=0}}
\caption{The ($s-r$) plane for the four-dimensional interacting dark energy model. Here we also show where the curves corresponding to quintessence and constant $w$ without interaction are located. The vertical dashed line corresponds to constant $w$ and the dotted line corresponds to quintessence. The three curves of Fig.~1 here are shown in the up left corner of the plot, close to $r=1$.}
\end{figure}

\begin{figure}
\centerline{\epsfig{figure=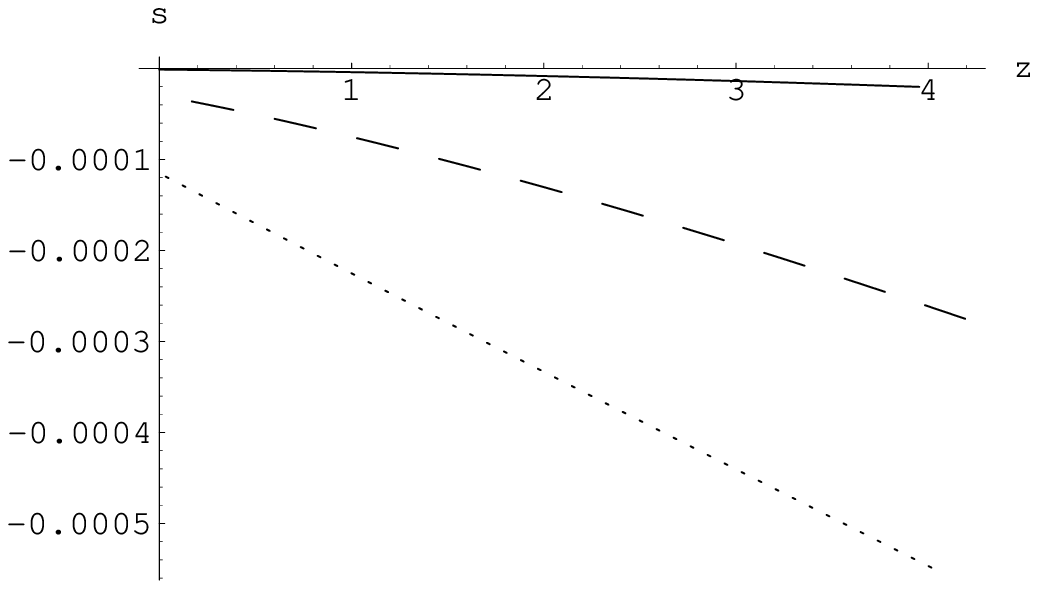,height=6cm,angle=0}}
\caption{Parameter $s$ as a function of red-shift $z$ for the brane model. The solid line corresponds to $\nu=1$, the dashed line corresponds to $\nu=1.1$ and the dotted line corresponds to $\nu=1.2$.}
\end{figure}

\begin{figure}
\centerline{\epsfig{figure=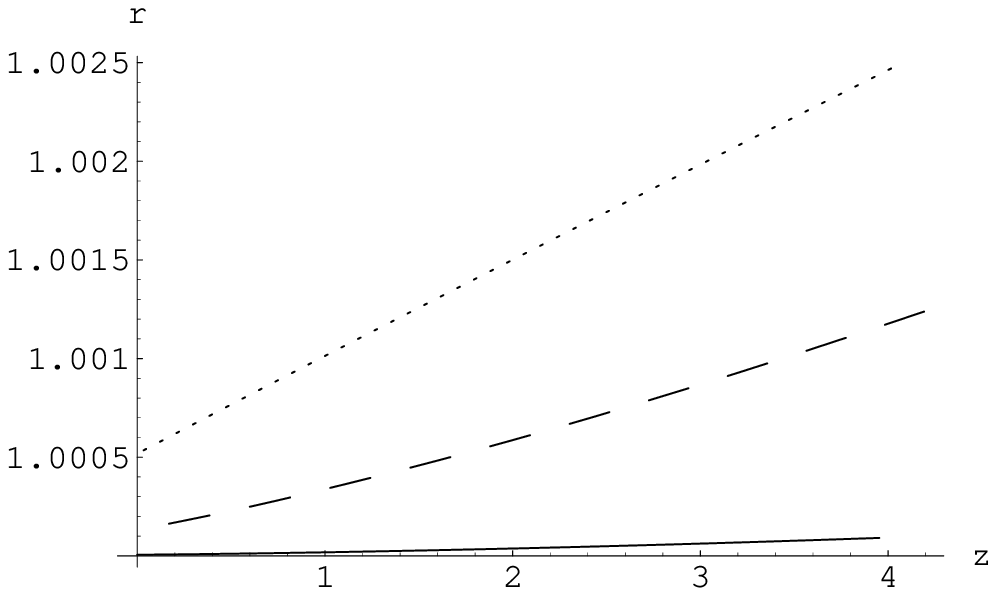,height=6cm,angle=0}}
\caption{Parameter $r$ as a function of red-shift $z$ for the brane model. The solid line corresponds to $\nu=1$, the dashed line corresponds to $\nu=1.1$ and the dotted line corresponds to $\nu=1.2$.}
\end{figure}

\end{document}